\documentclass[]{article}

\usepackage{amssymb}
\usepackage{graphicx}\usepackage{amsmath,amssymb}
\usepackage{color}
\usepackage[english]{babel}
\usepackage{xcolor}
\usepackage{srcltx}
\usepackage{greekbf}
\usepackage{nicefrac}
\usepackage{empheq}
\usepackage{enumerate}
\usepackage{wrapfig}
\usepackage{amsthm}

\usepackage{multicol}

\usepackage{mathtools}
\mathtoolsset{showonlyrefs}
    \mathtoolsset{centercolon}
\usepackage{mathrsfs}
\usepackage{epstopdf}

\allowdisplaybreaks

\DeclareMathAlphabet{\mathbf}{OT1}{cmr}{bx}{it}
\DeclareMathAlphabet{\mathssb}{OT1}{cmss}{bx}{n}
\DeclareMathAlphabet{\mathssn}{OT1}{cmss}{m}{n}
\DeclareMathAlphabet{\mathub}{OT1}{cmr}{b}{n}

\DeclareMathAlphabet{\mathpzc}{OT1}{pzc}%
                                 {m}{it}

\theoremstyle{definition}
\newtheorem{remark}{Remark}
\newcommand{\bydef}{\,\raise.050ex\hbox{\rm:}\kern-.025em\hbox{\rm=}\,}
\newcommand{\defby}{=\raise.075ex\hbox{\kern-.325em\hbox{\rm:}}\,}

\def\qed{\relax\ifmmode\hskip2em \Box\else\unskip\nobreak\hskip1em $\Box$\fi}
\newcommand {\Bc}  {\mathcal{B}}

\newcommand {\Kc}  {\mathcal{K}}
\newcommand {\Ic}  {\mathcal{I}}

\newcommand {\Lc}  {\mathcal{L}}

\newcommand {\Pc}  {\mathcal{P}}

\newcommand {\Sc}  {\mathcal{S}}
\newcommand {\Tc}  {\mathcal{T}}

\newcommand {\Wc}  {\mathcal{W}}

\newcommand {\ab} {\mathbf{a}}
\newcommand {\bb} {\mathbf{b}}
\newcommand {\cb} {\mathbf{c}}
\newcommand {\db} {\mathbf{d}}

\newcommand {\fb} {\mathbf{f}}

\newcommand {\hb} {\mathbf{h}}

\newcommand {\jb} {\mathbf{j}}
\newcommand {\kb} {\mathbf{k}}
\newcommand {\lb} {\mathbf{l}}
\newcommand {\mb} {\mathbf{m}}
\newcommand {\nb} {\mathbf{n}}
\newcommand {\pb} {\mathbf{p}}
\newcommand {\qb} {\mathbf{q}}
\newcommand {\rb} {\mathbf{r}}

\newcommand {\ub} {\mathbf{u}}
\newcommand {\vb} {\mathbf{v}}

\newcommand {\zb} {\mathbf{z}}
\newcommand {\Ab} {\mathbf{A}}

\newcommand {\Eb} {\mathbf{E}}
\newcommand {\Fb} {\mathbf{F}}

\newcommand {\Hb} {\mathbf{H}}

\newcommand {\Ib} {\mathbf{I}}

\newcommand {\Sb} {\mathbf{S}}

\newcommand {\Vb} {\mathbf{V}}

\newcommand {\alphab}     {\mathbf{\alpha}}

\newcommand {\etab}      {\mathbf{\eta}}

\newcommand {\vthetab}    {\mathbf{\vartheta}}

\newcommand {\xib}      {\mathbf{\xi}}

\newcommand {\varphib}   {\mathbf{\varphi}}

\newcommand {\omegab}    {\mathbf{\omega}}
\newcommand {\tr}[1]{\mbox{tr}\, #1}

\def\Div{\mathop{\hbox{Div}}}

\newcommand{\sym}{\mathop{\mathrm{sym}}}

\newcommand{\ov}{\overline}
\newcommand\jump[1]{[\![#1]\!]}

\begin{document}

\begin{center}
 {\bf \Large
A Beam Theory Consistent with

 \hspace{0.8cm}
 
 Three-Dimensional Thermo-Elasticity}
\end{center}
\medskip

\begin{center}
 {\large
Antonino Favata}
\end{center}

\begin{center}
 {\color{black}
Department of Civil, Environmental and Mechanical Engineering,\\ University of Trento\\ Via Mesiano, 77 -- 38123 Trento, Italy.\\antonino.favata@unitn.it
}
\end{center}

 \hspace{0.5cm}

\begin{abstract}
We  propose a model for thermo-elastic beams,  consistent with the  theory of linear three-dimensional thermo-elasticity and  deduced by a suitable  version of  the Principle of Virtual Powers. Dimensional reduction is achieved by postulating convenient \textit{a priori} representations for  mechanical and thermal displacements, the latter playing the role of an additional kinetic variable. Such representations are regarded as internal constraints, some involving the first, others the second gradient of deformation and thermal displacements; these constraints are maintained by reactive stresses and hyper-stresses of the type occurring in non-simple elastic materials of grade two, and by reactive entropy influxes and hyper-influxes.
\end{abstract}

 \hspace{0.4cm}


 \hspace{0.5cm}

\section{Introduction}
In continuum mechanics, the Principle of Virtual Powers is a standard tool to obtain all balance laws that apply to a given material class, be it the class of simple materials or, as suggested by Germain \cite{Ger}, the non-simple class of the so-called higher-gradient materials.\footnote{\color{black}Actually, in \cite{Auff,dellis}, it is argued that non-local and higher gradients continuum mechanics was considered already in Piola's works, by means of a suitable version of the Principle of Virtual Powers; the reader is referred to  \cite{Pio} and other works by the same author cited in \cite{Auff,dellis}. Therein it is possible to find other references to papers using the same spirit and methods as Piola to introduce
generalized stress tensors, e.g.  by Green and Rivlin \cite{GR,GR1,GR2,GR3}.} 
{\color{black}
In the same line of research, modern applications of the Principle can be found in \cite{al,dells1,pid}. 

}

Two uses of the Principle are nonstandard:
one is to derive   \textit{lower-dimensional} theories for thin structures, {\color{black} consistent with} a three-dimensional parent theory \cite{LPG,FPG,FPG1,primer2}; the other is to deduce the balance equations of a \textit{multiphysics} theory \cite{PPG}, that is, a theory that allows for composition of two or more material body structures, as is the case for the composition of mechanical and thermal structures considered in \cite{PPG}.

In this paper, we combine such nonstandard uses of the Principle to derive a thermo-elastic beam theory consistent with three-dimensional thermo-elasticity. Several models for beam thermo-mechanics have been proposed in the literature, some of them direct \cite{Bir,DSW,GN,GN1,K,ABE,Ies}, others induced from three-dimensional thermomechanics \cite{Jones_1966,Sim}. We here follow a different path from these valuable deductions; our goal is to achieve a consistency with three-dimensional thermo-elasticity, by using a generalization accommodating thermal phenomena of the  \textit{method of internal constraints} introduced in \cite{PPG_1989} in a purely mechanical context. According to that method,
 lower-dimensional structures are regarded as three-dimensional bodies, having a special shape and partitionability, and whose kinematics is restricted.\footnote{For a discussion of the different approaches adopted to model thin structures, see \cite{PPG_2008}. }

In our generalization to thermo-elasticity of the method of internal constraints,  an essential role is played by the notion of {\em thermal displacement}, a nonstandard thermal state variable firstly considered by Helmholtz.\footnote{A short account of the notion of thermal displacement, along with its history and use, can be found in \cite{PPG}.} In the spirit of the method, internal constraints have to be considered both on the mechanical and the thermal kinetic variables, constraints that are chosen so as to capture those characters of the lower-dimensional theory one considers most significant. 
It may happen -- as in the present case -- that a meaningful lower-dimensional kinetics (mechanical and thermal) emerge from some internal constraint formulated not only on the first gradient of the mechanical and/or thermal displacements, but also on second gradients. We are then led to consider \textit{non-simple thermo-elastic} materials.

The paper is organized as follows. In Section 2, we introduce our notation and define what we mean by a three-dimensional beam-like body. In Section 3, we recapitulate the extended version of the Principle of Virtual Power proposed in \cite{PPG} and we lay down the restrictions on the mechanical and thermal kinetic variables. Having done this, we find ourselves in a position to define the dynamical descriptors, to arrive at a one-dimensional version of the Principle of Virtual Power, and, finally, to obtain  the two balance laws of our beam theory, one for momentum the other for entropy. Section 4 is devoted to constitutive assumptions, consistent with a dissipation axiom leading to a reduced dissipation inequality. 

{\color{black} We observe that the balance equations obtained in Section 3 hold the same for a number of standard and non-standard beam theories, linear or non-linear, elastic or not. In Section 5,  we fix our attention on three-dimensional linear thermo-elasticity and wonder whether constitutive assumptions could be found such as to guarantee that our one-dimensional theory yields a sufficiently approximated description of the three-dimensional phenomenology. With this purpose,}
 we introduce the notion of nonsimple thermo-elastic materials and discuss the role of the modified dynamical descriptors, {\color{black}those which are constitutively specified and those that serves to maintain the internal  constraints.} 

The last two sections are devoted to exemplify the use of our theory. In Section 6, we show how to recover the standard equations of thermo-elastic beams, with some additional features; in particular, we consider the longitudinal-vibrations problem, and we discuss the relations between  one- and three-dimensional constitutive parameters. A second example,  bending vibrations, is considered in Section 7.

\section{Geometry}\label{geo}
\begin{figure}[h]
\centering
\includegraphics[scale=1.4]{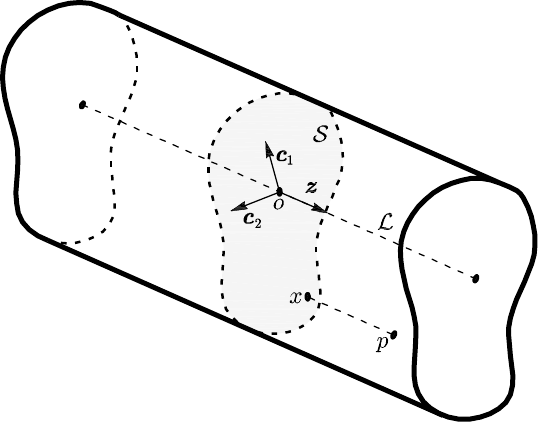}
\caption{A beam-like three-dimensional body.}
\label{fig1}
\end{figure}

We restrict our attention to beam-like solids in the form of right cylinders $\Bc$, of constant cross section $\Sc$ and length $l$, with $\Sc$ a flat, open, bounded and connected region with a smooth boundary $\partial\Sc$, whose diameter  is much smaller than $l$ (Figure \ref{fig1}).

For convenience, we introduce a Cartesian frame $\{o;\cb_1,\cb_2,\zb\}$ and we set:
\begin{equation}
\Bc\ni p=x+\zeta\zb, \quad x\in \Sc, \quad \zeta\in(-h,+h), \quad l=:2h;
\end{equation}
moreover, $x-o=x_\beta\, \cb_\beta$.\footnote{We use the convention that Latin and Greek indices have the range $\{1,2,3\}$ and $\{1,2\}$, respectively.} Region $\Bc$ can be identified point-wise with the set $\Sc\times(-h,+h)$; moreover,  we denote by
\begin{equation}
\Lc(x):=\{p\in\Bc\, | \, (p-x)\cdot\zb \in (-h,+h)\}
\end{equation}
the straight fiber of $\Bc$ through a point $x\in\Sc$. We fix once and for all the origin $o$ at the centroid of $\Bc$, so that $\Lc(0)$ is the axis, that we  shortly denote with $\Lc$.

\section{A non-standard use of the Principle of Virtual Power}
Given a material body  occupying a three-dimensional open and
bounded region $\Omega$ and a \textit{test} velocity field $\delta\ub$ over $\Omega$,
the \emph{internal virtual power expenditure} over a part $\Pc$ of
$\Omega$ associated with $\delta\ub$  is:
\[
\delta\Wc^{(i)}:=\int_\Pc\Sb\cdot\nabla\delta\ub,
\]
where  $\Sb$ denotes the stress field in $\Omega$; and,  the
\emph{external virtual power expenditure} over $\Pc$ is:
\[
\delta\Wc^{(e)}:=\int_\Pc\db\cdot \delta\ub+\int_{\partial\Pc}\cb\cdot\delta\ub,
\]
where $(\db,\cb)$ denote, respectively, the \emph{distance
force} for unit volume and the \emph{contact force} per unit area
exerted on $\Pc$ by its own complement with respect to $\Omega$
and by the environment of the latter. These representations of
power expenditures are those typical in the theory of the
so-called \emph{simple} material bodies. In that theory, a part is
customarily a subset of non-null volume of $\Omega$ (which makes
it for a part collection deemed sufficiently rich), and a
\emph{virtual velocity field} is a smooth vector field whose
support is a part; the \emph{Principle of Virtual Powers} (PVP)  is the
stipulation that
\begin{equation}\label{PVP}
\delta\Wc^{(i)}=\delta\Wc^{(e)}
\end{equation}
for all parts $\Pc$ of $\Omega$ and for all velocity fields $\delta\ub$ consistent with the admissible motions (see \cite{Ger,Mau_1980}).

In the form  \eqref{PVP}, the PVP is interpreted as a \emph{balance statement} for the internal and external fields  $\Sb$ and
$\db,\cb$; and as such, it concerns the  \textit{purely mechanical structure} of $\Omega$.

\subsection{Extended Powers, internal and external}

In \cite{PPG}, it is shown that the Virtual Power format can be so generalized as to deduce the balance laws of three-dimensional thermomechanics.  Together with the \textit{mechanical displacement} $\ub$, another kinetic variable is introduced, the \textit{thermal displacement} $\alpha$; both $\ub$ and $\alpha$ are required to be  smooth fields over the closure with respect to the product topology of the space-time cylinder $\Omega\times(0,T)$. Their time derivatives are the \textit{velocity} $\vb:=\dot{\ub}$ and the \textit{temperature}
 $$
 \vartheta:=\dot{\alpha}.
 $$ 
 For $\Pc$ a subbody of $\Omega$ and $\Tc=(t_i,t_f)$ a subinterval of $(0,T)$, the \textit{internal virtual power} is defined to be:
\begin{equation}\label{Wi}
\delta\Wc^{(i)}(\delta\ub,\delta\alpha):=\int_{\Pc\times\Tc}\Sb\cdot\nabla\delta\ub+ h\delta\alpha+\ov \hb\cdot\nabla\delta\alpha,
\end{equation}
where $ h$ and $\ov\hb$ are measures of the thermal interactions (of, respectively, zeroth and first order) of a material element of the subbody $\Pc$ with its immediate adjacencies;\footnote{One may ask why a zeroth order interaction has not been taken into account for the mechanical structure; as pointed out in \cite{PPG}, such a term would be cancelled by the requirement of translational invariance of $\delta\Wc^{(i)}$. } $h$ has to be interpreted as the \textit{internal dissipation} per unit temperature change, while $-\ov\hb=:\hb$  as a measure of \textit{entropy influx} at a point of an oriented surface of normal $\nb$ (see Remark 3 in \cite{PPG}).
The  `augmented' version of the external power is:
\begin{equation}\label{We}
\begin{aligned}
\delta\Wc^{(e)}(\delta\ub,\delta\alpha):=&\int_{\Pc\times\Tc}(\db\cdot\delta\ub+\pb\cdot\delta\dot{\ub}+d\delta\alpha+\eta\delta\dot{\alpha})+\int_{\partial\Pc\times  \Tc}(\cb\cdot\delta\ub+c\delta\alpha)+\int_{\Pc\times \partial I}\jump{\pb\cdot\delta\ub+\eta\delta\alpha},
\end{aligned}
\end{equation}
where
\begin{equation}
\begin{aligned}
\jump{\pb\cdot\delta\ub+\eta\delta\alpha}:=&\pb_f(x)\cdot\delta\ub(x,t_f)+\eta_f(x)\,\delta\alpha(x,t_f)+\pb_i(x)\cdot\delta\ub(x,t_i)+\eta_i(x)\,\delta\alpha(x,t_i),
\end{aligned}
\end{equation}
for all $x\in\Omega$; $\pb$ is the \textit{momentum} and $\eta$ the \textit{entropy}, both having specific sources $\db$ and $d$ at an interior point of $\Pc$, and specific \textit{fluxes} $\Sb\nb=\cb$ and $\ov \hb\cdot\nb=c$ at a boundary point of $\partial\Pc$. When the internal and external powers are taken as just detailed, the requirement \eqref{PVP} yields both momentum and entropy balance equations:
\begin{equation}
\dot{\pb}=\Div\Sb +\db, \quad \dot{\eta}=\Div\ov\hb -h +d.
\end{equation}
 We will use the latter  equation to describe heat conduction, a possibility that has been previously exploited in several frameworks; in particular,  it is worth mentioning the theory developed by Green and Naghdi, who introduced the  procedure of postulating an entropy balance in \cite{GN_1977} and proved in \cite{GN_1991} the consistency of this approach with the one stemmed by the energy balance;\footnote{For a revised exposition of Green-Naghdi theory, the reader is referred to \cite{BFPG}.} more recently,  a similar issue has been discussed in \cite{CBF,Fr}.

\subsection{Restrictions on kinetic variables}
To achieve dimensional reduction, it is necessary to adopt a restricted version of the three-dimensional statement of
PVP. The first
restriction  has to do with the \emph{special shape}
of the three-dimensional bodies we consider: they must be
beam-like, in the sense of Section \ref{geo}. The second
restriction concerns the choice of a \emph{special class of admissible body parts}: they all must have the same cross section of $\Bc$. Third and last, the \emph{class of virtual
velocities} is special, in that it is consistent with the representation of  admissible
displacements. The representations we choose are:
\begin{equation}\label{mecdisp}
\ub(x,\zeta,t)=\ab(\zeta,t)+\varphib(\zeta,t)\times(x-o),
\end{equation}
with
\begin{equation}\label{mecdisp1}
\ab(\zeta,t)=\vb_o(\zeta,t)+w(\zeta,t)\zb, \quad \vb_o(\zeta,t)\cdot\zb= \varphib(\zeta,t)\cdot\zb=0,
\end{equation}
for the mechanical displacement, and
\begin{equation}\label{therdisp}
\alpha(x,\zeta,t)=\alpha_o(\zeta,t)+ \alphab_1 (\zeta,t)\cdot(x-o), \quad \alphab_1(\zeta,t)\cdot\zb=0,
\end{equation}
for the thermal displacement. The representation \eqref{mecdisp}-\eqref{mecdisp1} is parametrized by the vector fields $\vb_o,\varphib$ over  $(-h,+h)\times(0,T)$; at a fixed time $\bar{t}$, the $\zeta$-cross section of abscissa  is subject to a translation $\vb_o(\zeta,\bar{t})+w(\zeta,\bar{t})\zb$ and a small rotation about an axis parallel to $\varphib(\zeta,\bar{t})$. {\color{black} Such a representation is typical of the so-called \textit{Timoshenko beam theory} \cite{Tim,Tim1}, which accounts for shear deformation and rotational inertia effects.  

}

 The \textit{Ansatz} \eqref{therdisp} on the thermal displacement is coherent with an analogous representation for the temperature distribution:
\begin{equation}\label{temp}
\vartheta(x,\zeta,t)=\vartheta_o(\zeta,t)+ \vthetab_1 (\zeta,t)\cdot(x-o), \quad \vthetab_1(\zeta,t)\cdot\zb=0;
\end{equation}
the representation \eqref{temp} is parametrized by the scalar field, the \textit{average} value $\vartheta_o$ -- a measure of the mean temperature over the cross section at abscissa $\zeta$ -- and by a \textit{deviation} $\vthetab_1$, a vector field identified with the  temperature gradient over the cross section. {\color{black} We notice that, if the \textit{Ansatz} \eqref{temp} is accepted, one contents with an approximated description, basically ruled by the value of the temperature at the intrados and the extrados of the beam.  As it has been pointed out by \cite{Jones_1966}: ``since the outer surface of the beam is insulated, it would not be unreasonable to assume a linear variation in temperature as well. For other boundary conditions, such as a given temperature on the surface, this assumption would not necessarily be warranted''. }
%

The kinetics of a thermoelastic beam is  specified by a list of twice continuously differentiable mappings
\begin{equation}\label{kinproc}
(\zeta,t)\mapsto \big(\ab(\zeta,t),\varphib(\zeta,t);\alpha_o(\zeta,t),\alphab_1(\zeta,t) \big);
\end{equation}
by time differentiation, a list 
\begin{equation}
(\zeta,t)\mapsto \big(\vb(\zeta,t),\omegab(\zeta,t);\vartheta_o(\zeta,t),\vthetab_1(\zeta,t)  \big)
\end{equation}
of \textit{realizable velocites} is obtained from each kinetic process \eqref{kinproc}.

\vspace{0.3cm}
\remark\label{EB} Instead of \textit{Timoshenko} beam,  the simpler \textit{Bernoulli-Navier} model obtains for
\begin{equation}\label{EBas}
\vb_o'-\varphib\times\zb=\mathbf{0},
\end{equation}
the requirement that cross section and axis keep mutually orthogonal in all admissible deformations.

\subsection{Internal power expenditure. Measures of stress, internal dissipation and entropy influx}
The first two restrictions on PVP we mentioned imply that a typical part of $\Omega\equiv\Bc$ can be identified with the Cartesian product $\Pc=\Sc\times\Ic$, with $\Ic=(a,b)$ an open and connected subset of $\Lc$. As to virtual velocities, we choose them of the form \eqref{mecdisp}-\eqref{mecdisp1} and \eqref{therdisp}:
\begin{equation}
\delta\ub(x,\zeta,t)=\delta\ab(\zeta,t)+\delta\varphib(\zeta,t)\times(x-o),
\end{equation} 
and 
\begin{equation}
\delta\alpha(x,\zeta,t)=\delta\alpha_o(\zeta,t)+ \delta\alphab_1 (\zeta,t)\cdot(x-o). 
\end{equation}
The displacements gradients are:
\begin{equation}
\begin{aligned}
&\nabla\delta \ub=\delta\ab'\otimes\zb+\big(\delta\varphib'\times(x-o)\big)\otimes\zb+\big(
\delta\varphib\times\cb_\beta\big)\otimes\cb_\beta,\\
& \nabla\delta\alpha=\big(\delta\alpha_o'+\delta\alphab_1'\cdot(x-o)\big)\,\zb+\delta\alphab_1,
\end{aligned}
\end{equation}
where we left the dependences on place and time tacit; here and henceforth, $(\cdot)'$ denotes the derivative with respect to $\zeta$; time derivative will be denoted by $(\cdot)^\cdot$. On having recourse to Fubini-Tonelli theorem, we obtain that
\begin{equation}
\begin{aligned}
\delta\Wc^{(i)}=&\int_{\Sc\times\Ic\times\Tc}\Sb\cdot\nabla\delta\ub+h\delta\alpha+\ov{\hb}\cdot\nabla\delta\alpha\\
=&\int_{\Ic\times\Tc}\Big(\delta \ab'\cdot\int_\Sc \Sb\zb +\delta\varphib\cdot\cb_\beta\times\int_{\Sc} \Sb\cb_\beta+\delta\varphib'\cdot\int_{\Sc}(x-o)\times\Sb\zb+ \delta\alpha_o\int_\Sc h+\delta\alphab_1\cdot\int_\Sc (x-o)h \\
+& \delta\alpha_o'\int_\Sc\ov{\hb}\cdot\zb+\delta\alphab_1'\cdot\int_\Sc (x-o)\ov{\hb}\cdot\zb+\delta\alphab_1\cdot\int_{\Sc}\ov{\hb} \Big).
\end{aligned}
\end{equation}
On making use of the following definitions:
\begin{equation}\label{defint}
\begin{aligned}
&\fb=\fb(\zeta,t):=\int_{\Sc}\Sb(x,\zeta,t)\zb,\\
&\mb=\mb (\zeta,t):=\int_\Sc(x-o)\times\Sb(x,\zeta)\zb,\\
& \xi_o=\xi_o(\zeta,t):=\int_\Sc h(x,\zeta,t),\\
&\xib_1= \xib_1(\zeta,t):=\int_\Sc (x-o)\,h(x,\zeta,t),\\
&\ov h_o= \ov h_o(\zeta,t):=\int_{\Sc}\ov{\hb}(x,\zeta,t)\cdot\zb,\\
&\ov{\hb}_1=\ov{\hb}_1(\zeta,t):=\int_\Sc (x-o)\,\ov{\hb}(x,\zeta,t)\cdot\zb,\\
&\ov\kb=\ov\kb(\zeta,t):=\int_\Sc\big(\ov{\hb}(x,\zeta,t)\cdot\cb_\beta\big)\,\cb_\beta
\end{aligned}
\end{equation}
the internal power expenditure can be given in the following form:
\begin{equation}
\begin{aligned}
\delta\Wc^{(i)}=&\int_{\Ic\times\Tc}\fb\cdot\delta\ab'+\mb\cdot\delta\varphib'+\zb\times\fb\cdot\delta\varphib+\int_{\Ic\times\Tc} \xi_o\,\delta\alpha_o+\xib_1\cdot\delta\alphab_1+\ov h_o\,\delta\alpha_o'+\ov\hb_1\cdot\delta\alphab_1'+\ov\kb\cdot\delta\alphab_1.\footnotemark
\end{aligned}
\end{equation} 
\footnotetext{We also made use of the algebraic identity $\cb_\beta\times\Ab\cb_\beta=\Ab\zb\times\zb$, where $\Ab$ is a symmetric tensor  \cite{primer2}.}

All quantities  in \eqref{defint} are  one-dimensional `dynamical' descriptors:  $\fb$ and $\mb$, the \textit{force} and the \textit{moment} vectors, are the \textit{stress measures}; $\xi_o$ and $\xib_1$ are the cross-sectional \textit{measures of internal dissipation}, \textit{average} and `\textit{deviational}', respectively;\footnote{We used the adjective `deviational' to recall that $\xib_1$ is power conjugated to the cross-sectional temperature deviation.} $-\ov h_o=:h_o$, $-\ov\kb=:\kb$ and $-\ov\hb_1=:\hb_1$ are the \textit{entropy influxes measures}, \textit{axial}, \textit{sectional} and \textit{deviational}, respectively. 

\subsection{External power expenditure. Applied loads, momenta, sources, and entropies}
As to the external power, we find:
\begin{equation}
\begin{aligned}
\delta\Wc^{(e)}=&\int_{\Sc\times\Ic\times\Tc}(\db\cdot\delta\ub+\pb\cdot\delta\dot{\ub}+d\delta\alpha+\eta\delta\dot{\alpha})+\int_{\partial(\Sc\times\Ic) \times\Tc}(\cb\cdot\delta\ub+c\delta\alpha)+\int_{\Sc\times\Ic\times\partial I}\jump{\pb\cdot\delta\ub+\eta\delta\alpha}=\\
&\int_{\Ic\times\Tc}\Big(\delta\ab\cdot\int_{\Sc}\db+\delta\varphib\cdot\int_{\Sc}(x-o)\times\db+\delta\dot{\ab}\cdot\int_{\Sc}\pb+\delta\dot{\varphib}\cdot\int_\Sc(x-o)\times\pb+\delta\alpha_o\int_\Sc d \\
+&\delta\alphab_1\cdot\int_\Sc(x-o)\, d+\delta\dot{\alpha}_o\int\eta+\delta\dot{\alphab}_1\cdot\int_{\Sc}(x-o)\,\eta  +\delta\ab\cdot\int_{\partial\Sc}\cb+\delta\varphib\cdot\int_{\partial\Sc}(x-o)\times\cb\\
+&\delta\alpha_o\int_{\partial\Sc} c+\delta\alphab_1\cdot\int_{\partial\Sc}(x-o)\, c\Big)+\int_{\Ic\times\partial\Tc}\Big[\!\!\Big[\Big(\delta\ab\cdot\int_\Sc\pb+\delta\varphib\cdot\int_\Sc(x-o)\times\pb\\
+&\delta\alpha_o\int_\Sc\eta+\delta\alphab_1\cdot\int_\Sc (x-o)\eta\Big)\Big]\!\!\Big].
\end{aligned}
\end{equation}
We are now in position to define over $\Lc$ applied loads, momenta, sources, and entropies, as  induced by three-dimensional loads $(\db,\cb)$, momentum $\pb$, specific source $d$, contact thermal interaction $c$, and entropy $\eta$. These are:
\begin{itemize}
\item the  \textit{force} and the \textit{couple}:
\begin{equation}
\begin{aligned}
& \bb_o=\bb_o(\zeta,t):=\int_\Sc\db(x,\zeta,t)+\int_{\partial\Sc}\cb(x,\zeta,t) ,\\
&\mb_o=\mb_o(\zeta,t):=\int_\Sc(x-o)\times\db(x,\zeta,t)+\int_{\partial\Sc}(x-o)\times\cb(x,\zeta,t);\\
\end{aligned}
\end{equation}
\item the \textit{linear} and \textit{rotational momentum}:
\begin{equation}
\begin{aligned}
&\lb_o=\lb_o(\zeta,t):=\int_\Sc\pb(x,\zeta,t),\\
&\rb_o=\rb_o(\zeta,t):=\int_\Sc(x-o)\times\pb(x,\zeta,t);\\
\end{aligned}
\end{equation}
\item the \textit{average} and \textit{deviational specific entropy sources}:
\begin{equation}
\begin{aligned}
& d_o=d_o(\zeta,t):=\int_\Sc d(x,\zeta,t)+\int_{\partial\Sc} c(x,\zeta,t),\\
& \db_1=d_1(\zeta,t):=\int_\Sc (x-o)\,d(x,\zeta,t)+\int_{\partial\Sc} (x-o)\,c(x,\zeta,t);\\
\end{aligned}
\end{equation}
\item the \textit{average} and \textit{deviational entropies}:
\begin{equation}
\begin{aligned}
& \eta_o=\eta_o(\zeta,t):=\int_\Sc \eta(x,\zeta,t),\\
& \etab_1=\etab_1(\zeta,t):=\int_\Sc (x-o)\,\eta(x,\zeta,t).\\
\end{aligned}
\end{equation}
\end{itemize}
With these, the internal power expenditure reads:
\begin{equation}
\begin{aligned}
\delta\Wc^{(e)}= &\int_{\Ic\times\Tc} \bb_o\cdot\delta\ab+ \mb_o\cdot\delta\varphib_o+\lb_o\cdot\delta\dot{\ab}+\rb_o\cdot\delta\dot{\varphib}+d_o\,\delta\alpha_o+\db_1\cdot\delta\alphab_1+\eta_o\,\delta\dot{\alpha}_o+\etab_1\cdot\dot{\alphab}_1\\
+&\int_{\Ic\times\partial\Tc}\jump{ \lb_o\cdot\delta\ab+\rb_o\cdot\delta\varphib+\eta_o\,\delta\alpha_o+\etab_1\cdot\alphab_1     }.
\end{aligned}
\end{equation} 

The one-dimensional \textit{entropy inflow} consists in the list $(h_o,\hb_1,\kb;d_o,\db_1)$, that we split in two:
\begin{itemize}
\item \textit{average}: $( h_o;d_o)$,
\item \textit{deviational}: $(\hb_1,\kb;\db_1)$.
\end{itemize}
We postulate that  as many \textit{heat influxes} exist: $(q_o;r_o)$, $(\qb_1,\jb;\rb_1)$ and, as usually done \cite{Mue_1971,PPG_2013,BFPG1}, we set the entropy inflow proportional to the heat inflow through the temperature as follows:
\begin{equation}\label{infl}
\begin{aligned}
& q_o=\vartheta_o\, h_o,  \\ 
& r_o=\vartheta_o\,d_o, \\ 
&\qb_1=\vartheta_1^\beta  h_1^\beta\cb_{\beta}, \qquad \vartheta_1^\beta:=\vthetab_1\cdot\cb_\beta, \quad  h_1^\beta:=\hb_1\cdot\cb_\beta,\\
&\jb=\vartheta_1^\beta  k^\beta\cb_{\beta}, \qquad  k^\beta:=\kb\cdot\cb_\beta,\\
& \rb_1=\vartheta_1\,\db_1.
\end{aligned}
\end{equation}

\subsection{Momentum and entropy balances}
The PVP, exploited for each virtual velocity defined over the closure of any subcylinder $\Ic\times\Tc$ of $\Lc\times(0,T)$ and such as to vanish at the ends of $\Tc$ itself, implies the momentum balances
\begin{equation}
\begin{aligned}\label{mombal}
& \dot{\lb}_o=\fb'+\bb_o,\\
& \dot{\rb}_o=\mb-\fb\times\zb+\mb_o,
\end{aligned}
\end{equation}
and the entropy balances
\begin{equation}\label{entrbal}
\begin{aligned}
& \dot{\eta}_o=\ov h_o'-\xi_o+d_o,\\
& \dot{\etab}_1=\ov \hb'-\ov\kb-\xib_1+\db_1,
\end{aligned}
\end{equation}
holding on $\Lc$, together with the initial conditions:
\begin{equation}\label{ic}
\lb_o(x,\zeta,t_i)=\lb_{oi}(x,\zeta), \quad \rb_o(x,\zeta,t_i)=\rb_{oi}(x,\zeta).
\end{equation}
As it always happens when a dimensional reduction is achieved, a loss of information has to be expected, which is partially mitigated by the appearance of richer dynamical descriptors and new balance equations.
{\color{black}It is worth noticing that, for whatever slenderness $d/l$, our procedure leads to the same balances equations; the choice of the \textit{Ans\"atze} \eqref{mecdisp} and \eqref{therdisp}, typical of those bodies inherently slender as beams are, prescribes the variations and then the format of the resulting one-dimensional balances equations we obtained. For these reasons we can say that \eqref{mecdisp} and \eqref{therdisp} are suitable for beams, as the resulting theory.}
{\color{black} While equations \eqref{mombal} are standard, the entropy balances \eqref{entrbal} are not: non-standard thermal descriptors appear and non-standard equations rule the phenomenon. 

In order to arrive at one or another beam model, be it elastic or not, some constitutive assumptions are necessary; in the next section, we show what of such assumptions are thermodynamically consistent. }

\section{Constitutive Assumptions}
Our constitutive assumptions stem for a \textit{dissipation axiom} (cf. \cite{PPG}), that is the requirement that, whatever kinetic process $(\zeta,t)\mapsto\big(\vb_o(\zeta,t),w(\zeta,t),\varphib(\zeta,t);$ $\alpha_o(\zeta,t),\alphab_1(\zeta,t) \big)$,
\begin{equation}\label{dissax}
\xi_o\dot{\alpha_o}\leq 0, \quad \xib_1\cdot\dot{\alphab}_1\leq 0,
\end{equation}
over the space-time cylinder $\Lc\times(0,T)$.

Now, let the specific \textit{internal action} per unit length $\phi$ be defined by
\begin{equation}\label{intac}
\begin{aligned}
\dot{\phi}:=&\fb\cdot\vb'+\mb\cdot\omegab'+\zb\times\fb\cdot\omegab-(\lb_o\cdot\dot\vb+\rb_o\cdot\dot{\omegab})+\xi_o\vartheta_o+\xib_1\cdot\vthetab_1+\ov h_o\vartheta_o'+\ov\hb_1\cdot\vartheta_1'+\ov\kb\cdot\vartheta_1\\
-&(\eta_o\dot{\vartheta}_o+\etab_1\cdot\dot{\vthetab}_1),
\end{aligned}
\end{equation}
and let $\kappa$ be the specific \textit{kinetic energy} per unit length:
\begin{equation}\label{kappa}
\kappa:=\frac{1}{2}(\lb_o\cdot\vb+\rb_o\cdot\omegab),
\end{equation}
so that
\begin{equation}
\Kc(\Ic)=\frac{1}{2}\int_{\Ic\times\Sc}\pb\cdot\dot{\ub}=\frac{1}{2}\int_{\Ic} \lb_o\cdot\vb+\rb_o\cdot\omegab=\frac{1}{2}\int_{\Ic}\kappa
\end{equation}
is the kinetic energy of the part $\Ic$. In classical mechanics, the link between the time rate of the kinetic energy and the power $\Pi^{in}$ expended by the inertia forces $\db^{in}=-\dot{\pb}$ is such that $\dot{\Kc}+\Pi^{in}=0$ (see \cite{PPG_1997}, \cite{FPGT}), and then
\begin{equation}
\dot{\kappa}=\dot{\lb_o}\cdot\vb +\dot{\rb_o}\cdot\omegab.
\end{equation}
Let us introduce the specific \textit{energy} 
\begin{equation}
\tau:=\phi+\lb_o\cdot\vb+\rb_o\cdot\omegab+\eta_o\vartheta_o+\etab_1\cdot\vthetab_1=\phi+2\kappa+\eta_o\vartheta_o+\etab_1\cdot\vthetab_1
\end{equation}
and set
\begin{equation}\label{freene}
\epsilon=\tau-\kappa, \quad \phi=\psi-\kappa,
\end{equation}
with $\epsilon$ the specific \textit{internal energy} per unit length and $\psi$ the specific \textit{Helmholtz free energy} per unit length, whence
\begin{equation}
\psi=\epsilon-(\eta_o\vartheta_o+\etab_1\cdot\vthetab_1).
\end{equation}

The dissipation axiom \eqref{dissax}, combined with \eqref{freene}, \eqref{kappa}, \eqref{intac} yields the \textit{dissipation inequality}
\begin{equation}
\begin{aligned}
\dot{\psi}\leq&-(\eta_o\dot\vartheta_o+\etab_1\cdot\dot\vthetab_1)+\ov h_o\vartheta_o'+\ov\hb_1\cdot\vthetab_1'+\ov\kb\cdot\vthetab_1+\fb\cdot\vb'+\mb\cdot\omegab'+\zb\times\fb\cdot\omegab,
\end{aligned}
\end{equation}
or, rather better,
\begin{equation}\label{diseg}
\begin{aligned}
\dot{\psi}\leq&-(\eta_o\dot\vartheta_o+\etab_1\cdot\dot\vthetab_1)- h_o\vartheta_o'- \hb_1\cdot\vthetab_1'- \kb\cdot\vthetab_1+\fb\cdot\vb'+\mb\cdot\omegab'+\zb\times\fb\cdot\omegab.
\end{aligned}
\end{equation}

\vspace{0.3cm}
\remark All the quantities above defined can be split into a linear and rotational part, denoted with $(\cdot)_o$ and $(\cdot)_1$, respectively:
\begin{itemize}
\item internal action
\begin{equation}
\begin{aligned}
\dot\phi_o:=&\fb\cdot\vb'-\lb_o\cdot\dot\vb+\xi_o\vartheta_o+\ov h_o\vartheta_o'+-\eta_o\dot{\vartheta}_o,\\
\dot{\phi}_1:=&\mb\cdot\omegab'+\zb\times\fb\cdot\omegab-\rb_o\cdot\dot{\omegab}+\xib_1\cdot\vthetab_1+\ov\hb_1\cdot\vartheta_1'+\ov\kb\cdot\vartheta_1-\etab_1\cdot\dot{\vthetab}_1;
\end{aligned}
\end{equation}
\item kinetic energy
\begin{equation}
\begin{aligned}
\kappa_o:=\frac{1}{2}\lb_o\cdot\vb, \qquad \kappa_1:=\frac{1}{2}\rb_o\cdot\omegab
\end{aligned}
\end{equation}
\item energy
\begin{equation}
\begin{aligned}
&\tau_o:=\phi_o+\lb_o\cdot\vb+\eta_o\vartheta_o=\phi_o+2\kappa_o+\eta_o\vartheta_o,\\
&\tau_1:=\phi_1+\rb_o\cdot\omegab+\etab_1\cdot\vthetab_1=\phi_1+2\kappa_1+\etab_1\cdot\vthetab_1;
\end{aligned}
\end{equation}
\item internal energy
\begin{equation}
\begin{aligned}
\epsilon_o:=\tau_o-\kappa_o,\qquad \epsilon_1:=\tau_1-\kappa_1;
\end{aligned}
\end{equation}
\item Helmholtz free energy
\begin{equation}
\begin{aligned}
\psi_o:=\phi_o+\kappa_o, \qquad \psi_1:=\phi_1+\kappa_1.
\end{aligned}
\end{equation}
\end{itemize}
With these definitions, the dissipation inequality \eqref{diseg} is equivalent to the following two inequalities:
\begin{equation}\label{diss}
\begin{aligned}
&\dot{\psi}_o\leq-\eta_o\dot\vartheta_o- h_o\vartheta_o'+\fb\cdot\vb',\\
&\dot{\psi}_1\leq -\etab_1\cdot\dot\vthetab_1-\hb_1\cdot\vthetab_1'-\kb\cdot\vthetab_1+\mb\cdot\omegab'+\zb\times\fb\cdot\omegab.
\end{aligned}
\end{equation}

{\color{black} The balance equations  \eqref{mombal}--\eqref{entrbal}, together with one or another set of constitutive assumptions coherent with  dissipation inequalities \eqref{diss}, lead to a number of standard and non-standard beams theory, linear or non-linear, elastic or not. In the following, we fix our attention to linear thermo-elasticity and wonder if, on adopting constitutive assumptions consistent with the three-dimensional version of that theory, the predictions we obtained for our one-dimensional theory are `consistent' and in what sense `approximating'.}

\section{Reactive stresses and reactive entropy influxes}

Once a constitutive choice has been made and a solution of the problem governed by equations \eqref{mombal}-\eqref{entrbal} has been found, one wonders how accurately it approximates the three-dimensional  \textit{thermo-elastic state} $\{\ub,\Eb,\Sb;\alpha,\hb\}^{\rm E}$ that solves \textit{exactly} the parent three-dimensional problem that our lower-dimensional model aims to approximate. As to the mechanical and thermal displacements, we approximate  $\ub^{\rm E}$ and $\alpha^{\rm E}$ by inserting the solution $(\ab,\varphib;\alpha_o,\alphab_1)$ into the representations \eqref{mecdisp} and \eqref{therdisp}; consequently, we obtain the approximation for $\Eb^{\rm E}$. As to the stress field $\Sb^{\rm E}$ and the entropy influx $\hb^{\rm E}$, there are various approximations, scrutinized, for the stress, in \cite{LPG}. 

Firstly, an \textit{active} (i.e. constitutively determined) stress field $\Sb^{\rm A}$ and an active entropy influx field $\hb^{\rm A}$ can be considered. As done in \cite{PPG_1989,LPG} for the stress, a better approximation of $\Sb^{\rm E}$ and $\hb^{\rm E}$ can be obtained  by adding a \textit{reactive} stress and entropy influx fields, regarded as consequences of the kinematical restrictions. The reactive stresses are specified by the constitutive requirement of doing no work in any admissible deformation (see Section 30 of \cite{Tr}):
\begin{equation}
\Sb^{\rm R}\cdot\nabla \ub=0
\end{equation}
for grade-1 materials of Cauchy,
\begin{equation}\label{vincu}
\Sb^{\rm R}\cdot\nabla \ub+\boldsymbol{\mathsf{S}}^{\rm R}\cdot\nabla\nabla\ub=0,
\end{equation}
(cf. \cite{LPG}) for materials grade-2; $\Sb^{\rm R}$ and $\boldsymbol{\mathsf{S}}^{\rm R}$ are a \textit{reactive stress} and a \textit{reactive hyper-stress}, respectively (see the next subsection). Similarly, in our extended framework, we assume that reactive entropy influxes are specified by the analogous requirement:
\begin{equation}\label{vinca}
\hb^{\rm R}\cdot\nabla\alpha+\Hb^{\rm R}\cdot\nabla\nabla\alpha=0,
\end{equation}
where $\hb^{\rm R}$ and $\Hb^{\rm R}$ are a \textit{reactive entropy influx} and a\textit{ reactive entropy hyper-influx}. We will make this statement more precise in the next subsection.

\subsection{Simple and nonsimple reactive stresses and entropy influxes}\label{reacs}

In classical constitutive theories, the stress of a body particle is decided by the deformation history of a neighbourhood of that particle;  in a \textit{simple} material the stress is a function of the history of the deformation gradient $\Fb$, the \textit{first-order approximation} of the deformation; in a simple and \textit{elastic} material, the stress depends on the present value of $\Fb$. In general, the deformation can be better approximated considering deformation gradients of order $N>1$, leading to \textit{nonsimple elastic materials of grade} $N>1$. For this kind of materials, the concept of stress has to be reconsidered, by introducing stresses of high order. We focus on materials of grade 2, for which the standard stress $\Sb$ has to be accompanied by the \textit{hyper-stress} $\boldsymbol{\mathsf{S}}$, a third order tensor, in that the internal virtual power becomes:
\begin{equation}
\int_\Pc \Sb\cdot\nabla\delta\ub+\boldsymbol{\mathsf{S}}\cdot\nabla\nabla\delta\ub.
\end{equation}
So far, we were concerned with the purely mechanical structure of the body; within our framework, it is not unreasonable to think of a heat conductor, for which it is possible to give a convenient definition of \textit{non-simple conductor material} of grade $M>1$: a material whose entropy influxes depend on the $M$-th gradient of the thermal displacement;  together with the notion of \textit{hyper-stress}, an \textit{entropy hyper-influx} comes out. Accordingly, thermo-elastic grade-$(N-M)$ materials are those for which the stresses and entropy influxes depend on the $N$-th and $M$-th gradient of the mechanical and thermal displacements, respectively; in principle, there is no reason to assume $N=M$. We here consider the case when $N=M=2$, whose
 corresponding extended internal virtual power reads:
\begin{equation}
\int_{\Pc\times\Tc}\Sb\cdot\nabla\delta\ub+\boldsymbol{\mathsf{S}}\cdot\nabla\nabla\delta\ub+h\delta\alpha+\ov\hb\cdot\nabla\delta\alpha+\ov\Hb\cdot\nabla\nabla\delta\alpha,
\end{equation}
where $-\ov\Hb=:\Hb$ is the entropy hyper-influx, a second order tensor. 

 It is  not difficult to see that the \textit{modified stress}
\begin{equation}
\widetilde{\Sb}:=\Sb-\Div\boldsymbol{\mathsf{S}},
\end{equation}
and the \textit{modified entropy influx}
\begin{equation}
\widetilde{\hb}:=\hb-\Div \Hb
\end{equation}
enter the equilibrium and the entropy balance equations. In general, each of $\Sb$, $\boldsymbol{\mathsf{S}}$, $\hb$ and $\Hb$ consists of an \textit{active} (that is constitutively determined) and a \textit{reactive} part, arising in presence of internal constraints. If they are expressed by scalar equations
\begin{equation}
\gamma\big(\Eb(\ub) \big)=0, \quad \Gamma\big(\nabla\nabla\ub\big)=0,\quad \sigma\big(\nabla\alpha\big)=0, \quad \Sigma(\nabla\nabla\alpha)=0,
\end{equation}
then conditions \eqref{vincu}-\eqref{vinca} yield the following representation for the associated reactive stress, hyper-stress, entropy influx and entropy hyper-influx:
\begin{equation}\label{reactions}
\begin{aligned}
&\Sb^{\rm R}=S^{\rm R}\big(\partial_\Eb\gamma\big), \quad  \boldsymbol{\mathsf{S}}^{\rm R}=\mathsf{S}^{\rm R}\,\big(\partial_{\nabla\nabla\ub}\Gamma \big), \\
& \hb^{\rm R}= h^{\rm R}\big( \partial_{\nabla\alpha}\sigma\big), \quad \Hb^{\rm R}= H^{\rm R}\big( \partial_{\nabla\nabla\alpha}\Sigma\big),
\end{aligned}
\end{equation}
with $S^{\rm R},\mathsf{S}^{\rm R},h^{\rm R},H^{\rm R}$ four arbitrary scalars.

{\color{black} It is possible to show \cite{LPG} that the kinematic assumptions \eqref{mecdisp}-\eqref{mecdisp1} are the general solutions the following linear system of partial differential equations; 
 \begin{equation}\label{intconstr}
 \Eb\cdot\cb_\beta\otimes\cb_\gamma=0, \quad \Eb,_\gamma \cdot\zb\otimes\cb_\beta=0,
 \end{equation}
 where
 \begin{equation}
 \Eb:=\sym\nabla\ub=\frac{1}{2}(\nabla\ub+\nabla\ub^T)
 \end{equation}
 is the infinitesimal strain tensor. Equations \eqref{intconstr} are regarded as \textit{internal constraints}, because they imply that not all states of strain are admissible; in particular, they imply that the cross section fibers neither lengthen nor change their mutual angle and that, given a fiber on the cross section, the change in angle between those fiber and an axial fiber has a constant value.
 Analogously, it is not difficult to see that the condition \eqref{therdisp}  is the solution of the following internal constraints:
 \begin{equation}\label{condth}
 \nabla\nabla\alpha\cdot\cb_\beta\otimes\cb_\gamma=0.
 \end{equation}
 The four scalar conditions equivalent to \eqref{condth} are in fact:
 \begin{equation}\label{pdes}
 \alpha,_{11}(x_1,x_2,\zeta)=\alpha,_{22}(x_1,x_2,\zeta)=\alpha,_{12}(x_1,x_2,\zeta)=0,
 \end{equation}
 and an easy computation shows that the field solving the three first order PDEs \eqref{pdes} is \eqref{therdisp}.
The simple constraints \eqref{intconstr}$_1$ produce an admissible reactive stress having the form }
\begin{equation}
\begin{aligned}
\Sb^{\rm R}=\sum_{i=1}^{3}S^{\rm R}_i\,\Vb_i, \quad &\Vb_1:=\cb_1\otimes\cb_1, \Vb_2:=\cb_2\otimes\cb_2,\\ &\Vb_3:=\cb_1\otimes\cb_2+\cb_2\otimes\cb_1.
\end{aligned}
\end{equation}
Moreover, when the second order internal constraints \eqref{intconstr}$_2$ hold, the constraint equations are:
\begin{equation}
\Gamma_{\beta\delta}= \nabla\nabla\ub\cdot\Big(\zb\otimes\sym (\cb_\beta\otimes\cb_\delta) +\cb_\beta\otimes \sym (\zb\otimes\cb_\delta) \Big)=0;
\end{equation}
the representation \eqref{reactions}$_2$ becomes
\begin{equation}
\widetilde{\boldsymbol{\mathsf{S}}}^{\rm R}=\Sigma_{\beta\delta}\Big(\zb\otimes\sym (\cb_\beta\otimes\cb_\delta) +\cb_\beta\otimes \sym (\zb\otimes\cb_\delta) \Big).
\end{equation}
From the simple constraint \eqref{intconstr}$_1$ follows that
\begin{equation}
\gamma_{\beta\delta,k}(\nabla\ub)=E_{\beta\delta,k}=0, \quad k=1,2,3, 
\end{equation}
whose corresponding non simple constraint is
\begin{equation}
\nabla\nabla\ub\cdot\Big(\cb_\beta\otimes\sym (\cb_\delta\otimes\cb_k)+\cb_\delta\otimes\sym (\cb_\beta\otimes\cb_k)    \Big)=0, 
\end{equation}
where we mean $\cb_3\equiv\zb$; the corresponding hyper-stress is then
\begin{equation}
\overline{\boldsymbol{\mathsf{S}}}^{\rm R}=\tau_{\beta\delta k}\Big(\cb_\beta\otimes\sym (\cb_\delta\otimes\cb_k)+\cb_\delta\otimes\sym (\cb_\beta\otimes\cb_k)    \Big),  \; \tau_{\beta\delta k}=\tau_{\delta\beta k}.
\end{equation}
All in all, the reactive hyper-stress has the form:
\begin{equation}
\begin{aligned}
\boldsymbol{\mathsf{S}}^{\rm R}&=\widetilde{\boldsymbol{\mathsf{S}}}^{\rm R}+\overline{\boldsymbol{\mathsf{S}}}^{\rm R}=\\
&=(\tau_{\beta\delta\mu}+\tau_{\mu\beta\delta})\,\cb_\beta\otimes\cb_\delta\otimes\cb_\mu+\frac{1}{2}(\Sigma_{\beta\delta}+\Sigma_{\delta\beta})\,\zb\otimes\cb_\beta\otimes\cb_\delta+(\Sigma_{\beta\delta}+2\tau_{\beta\delta 3})\,\cb_\beta\otimes\sym (\cb_\delta\otimes\zb)
\end{aligned}
\end{equation}
(cf. \cite{LPG}). 

 Concerning the thermal structure, we do not have any constraint on $\nabla\alpha$, so that $\hb^{\rm R}=\mathbf{0}$; as to $\nabla\nabla\alpha$, the constraint equations are:
\begin{equation}
\Sigma_{\beta\delta} \big(\nabla\nabla\alpha\big)=\nabla\nabla\alpha\cdot\cb_\beta\otimes \cb_\delta=0
\end{equation}
(cf. \eqref{condth}); the representation \eqref{reactions}$_4$ becomes:
\begin{equation}
\Hb^{\rm R}=\sum_{i=1}^{3} H^{\rm R}_{i}\,\Vb_i.
\end{equation}
We conclude noticing that the beam we are considering can be then regarded as a grade-$(2-2)$ thermo-elastic body, in which the hyper-stress and entropy hyper-influx are completely reactive; the stress is partially active and partially reactive; the entropy influx is entirely active.

\vspace{0.3cm}
\remark 
To our best knowledge, this is the first time that the notion of non-simple conductor material of grade $M>1$ is introduced, together with the modified entropy influx. Nevertheless, `augmented' constitutive choices for the entropy influx have been proposed for instance by Green and Naghdi in their so-called Type II and Type III theories. In the case of the latter \cite{GN_1992}, they propose to take
\begin{equation}\label{heatflux3en}
\widehat{\boldsymbol{h}}(\alpha,\vartheta,\nabla\alpha,\nabla\vartheta)=-\kappa\,\vartheta^{-1}\,\nabla\vartheta-\big(\kappa^\star+\kappa^{\star\star}\big)\,\vartheta^{-1}\,\nabla\alpha,
\end{equation}
where $\kappa^\star$ and $\kappa^{\star\star}$ are two new conductivity moduli; the corresponding assumption for the heat influx is:
\begin{equation}\label{heatflux3}
\widehat{\boldsymbol{q}}(\alpha,\vartheta,\nabla\alpha,\nabla\vartheta)=-\kappa\,\nabla\vartheta-\big(\kappa^\star+\kappa^{\star\star}\big)\,\nabla\alpha.
\end{equation}
The first term of RHS of \eqref{heatflux3en}  is the standard  Fourier assumption, leading to the classical heat conduction. We can interpret \eqref{heatflux3en} as the expression of a grade-2 modified entropy influx $\widetilde{\hb}$, for which the following constitutive assumptions have been made:
\begin{equation}
\hb=\widehat{\hb}(\alpha,\vartheta,\nabla\alpha,\nabla\vartheta)=-\kappa\,\vartheta^{-1}\,\nabla\vartheta,
\end{equation}
for the entropy influx, and
\begin{equation}\label{H}
\Div\Hb =\Div\widehat{\Hb}(\alpha,\vartheta,\nabla\alpha,\nabla\vartheta)=(\kappa^\star+\kappa^{\star\star})\vartheta^{-1}\nabla\alpha, 
\end{equation}
for the entropy hyper-influx.

\section{A first example: longitudinal vibrations}
In this  section, we consider an elementary problem, which allow us to recover the standard equations of thermoelastic beams. First of all, here and henceforth, we restrict our attention to Euler-Bernoulli (E-B) beams (see Remark \ref{EB}); moreover, we strengthen the kinetic restrictions on $\alpha$, assuming that the following constraint holds:
\begin{equation}\label{constra}
\sigma(\nabla\alpha)=\nabla\alpha\cdot\cb_\beta=0.
\end{equation}  
According to representation \eqref{reactions}, this further assumption produces reactive entropy influxes 
\begin{equation}
\hb^{\rm R}_\beta= h^{\rm R}_\beta\cb_\beta,
\end{equation}
while the constitutively determined part of $\hb$ is just the one directed as $\zb$. Moreover, \eqref{constra} implies that $\alphab_1=\mathbf{0}$.

As to the inertial forces, considered for the three-dimensional beam-like body $\Bc$, we make the usual assumptions, that is:
\begin{equation}
\db^{in}(x,\zeta,t)=-\dot\pb(x,\zeta,t)=-\rho(x,\zeta)\,\ddot{\ub}(x,\zeta,t),
\end{equation}
where $\ub$ is consistent with \eqref{mecdisp}-\eqref{mecdisp1} and \eqref{EBas}, and  $\rho(x,\zeta)$ is the mass density per unit volume.
Both sources and non inertial forces are assumed to be null.

It is not difficult to see that balance equations \eqref{mombal} and \eqref{entrbal} reduce to:
\begin{equation}\label{sys}
\left\{ \begin{array}{ll}
-\rho_o \ddot{w}&=N',\\
 \dot{\eta}_o&=- h_o'-\xi_o
\end{array} \right.
\end{equation} 
where $N:=\fb\cdot\zb$ is the \textit{normal force} and 
$$
\rho_o=\rho_o(\zeta):=\int_\Sc\rho(x,\zeta)
$$
is the mass density per unit length. We aim to find a coupled system of PDEs describing mechanical effects and heat conduction. Constitutive assumptions stem from the use \textit{\`a la} Coleman-Noll \cite{CN} of inequality \eqref{diss}, which takes the form:
\begin{equation}\label{diso}
\dot{\psi}_o\leq-\eta_o\dot\vartheta_o- h_o\vartheta_o'+N\dot\varepsilon, 
\end{equation}
with $\varepsilon:=w'$ the \textit{axial strain} of the beam. As customary, we assume that the quantities in need of a constitutive prescription depend on one and the same list of variables:
\begin{equation}
\begin{aligned}
&\psi_o=\widehat{\psi}_o(\varepsilon,\vartheta_o, \vartheta_o'), \quad \eta_o=\widehat{\eta}_o(\varepsilon,\vartheta_o, \vartheta_o'), \\ & N=\widehat{N}(\varepsilon,\vartheta_o, \vartheta_o').
\end{aligned}
\end{equation} 
Thus, inequality \eqref{diso} reads:
\begin{equation}\label{CNo}
(\partial_{\vartheta_o}\psi_o+\eta_o)\,\dot{\vartheta}_o+(\partial_\varepsilon\psi_o-N)\,\dot{\varepsilon}+\partial_{\vartheta_o'}\psi_o\,\dot\vartheta_o'+ h_o\vartheta_o'\leq 0 ;
\end{equation}
we now require that \eqref{CNo}  be satisfied whatever the local continuation of any conceivable process, that is, whatever $(\dot{\varepsilon}, \dot\vartheta_o, \dot{\vartheta}_o')$ at whatever state $(\varepsilon, \vartheta_o,\vartheta_o')$. This requirement is satisfied if and only if
\begin{equation}\label{constase}
\begin{aligned}
&\widehat{\psi}_o \;{\rm is \; independent \; of \; \vartheta_o'}, \quad \widehat{\eta}_o(\varepsilon,\vartheta_o)=-\partial_{\vartheta_o}\widehat{\psi}_o(\varepsilon,\vartheta_o), \\ & \widehat{N}(\varepsilon,\vartheta_o)=\partial_\varepsilon\widehat{\psi}_o(\varepsilon,\vartheta_o),
\end{aligned}
\end{equation}
and moreover, 
$$
\widehat{h}_o(\varepsilon,\vartheta_o,\vartheta_o')\,\vartheta_o'\leq 0, 
$$
for all $\varepsilon$, $\vartheta_o$ and $\vartheta_o'$; on using \eqref{infl}$_1$, this latter condition is equivalent to 
\begin{equation}\label{qt}
\widehat{q}_o(\varepsilon,\vartheta_o,\vartheta_o')\,\vartheta_o'\leq 0.
\end{equation}
Condition \eqref{qt} implies that 
\begin{equation}\label{four}
\widehat{q}_o(\varepsilon,\vartheta_o,\vartheta_o')=-\widehat{\chi}(\varepsilon,\vartheta_o,\vartheta_o')\,\vartheta_o',
\end{equation}
where $\widehat{\chi}$ is the conductivity mapping; as is usually done, we assume that $\widehat{\chi}$ has a constant value, whence
\begin{equation}
h_o=-\chi\vartheta_o^{-1}\,\vartheta_o'=-\chi(\log\vartheta_o)'.
\end{equation}
We now choose the free energy as follows:
\begin{equation}\label{fren}
\widehat{\psi}_o(\varepsilon,\vartheta_o)=\frac{1}{2}s_E\,\varepsilon^2+m\,\varepsilon(\vartheta_o-\bar{\vartheta}_o)-c_o\vartheta_o\log\vartheta_o,
\end{equation}
where $s_E$ is the \textit{extensional stiffness} of the beam, $m$ the \textit{stress-temperature modulus}, $\bar{\vartheta}_o$ a prescribed (constant) value of the mean temperature over the cross section, and $c_o$ the \textit{heat capacity} per unit length; we will discuss the relation between these moduli and their correspondent three-dimensional version in Section \ref{constpar}. With this choice, \eqref{constase} implies:
\begin{equation}\label{constase1}
\widehat{\eta}(\varepsilon,\vartheta_o)=c_o(1+\log\vartheta_o)-m\varepsilon,  \quad  \widehat{N}(\varepsilon,\vartheta_o)=s_E\varepsilon+m(\vartheta_o-\bar{\vartheta}_o).
\end{equation}
Note that, if $s_E\neq 0$, as we request, then  \eqref{constase1}$_2$ can be inverted to yield:
\begin{equation}
\varepsilon=\frac{N}{s_E}+\delta_o(\vartheta_o-\bar{\vartheta}_o),
\end{equation}
where
$$
\delta_o:=-\frac{m}{s_E}
$$
is the \textit{coefficient of thermal dilation} (cf. eq. (8.4), Section 8 in \cite{Carl}). 
As to the dissipation $\xi_o$, as in three-dimensional thermo-elasticity \cite{CBF,Fr}, we set:
\begin{equation}
\xi_o=q_o\vartheta_o'=-\chi\,(\vartheta_o')^2.
\end{equation}

All in all, the system \eqref{sys} becomes:
\begin{equation}
\left\{ \begin{array}{ll}
&-\dfrac{\rho_o}{s_E} \ddot{w}= w''-\delta_o \vartheta_o',\\ [0.2cm]
& c_o(\log\vartheta_o)^\cdot-\chi(\log\vartheta_o)''=-\delta_o\,s_E\dot{w}'+\chi(\vartheta_o')^2.
\end{array} \right.
\end{equation} 
If dissipation is neglected because it is a quadratic quantity, we obtain:
\begin{equation}\label{syst}
\left\{ \begin{array}{ll}
&-\dfrac{\rho_o}{s_E} \ddot{w}= w''-\delta_o \vartheta_o',\\ [0.2cm]
& c_o(\log\vartheta_o)^\cdot-\chi(\log\vartheta_o)''=-\delta_o\,s_E\dot{w}'.
\end{array} \right.
\end{equation} 
If we make borrow from \cite{CBF,Fr} the small perturbation assumption  $\vartheta_o=\tilde{\vartheta}_o+\theta(\zeta,t)$, where $\theta$ is the small perturbation, the system \eqref{syst} becomes:
\begin{equation}\label{odes}
\left\{ \begin{array}{ll}
&-\dfrac{\rho_o}{s_E} \ddot{w}= w''-\delta_o \theta_o',\\ [0.2cm]
& c_o\dot{\theta}-\chi\theta''=-\tilde{\vartheta}_o\,\delta_o\,s_E\,\dot{w}'.
\end{array} \right.
\end{equation} 
The first equation is  well-known in technical beam theory; the second one is similar to the classical heat conduction equation, where the time rate of axial deformation $\dot{\varepsilon}=\dot{w}'$ plays the role of a heat source.

\subsection{One- and three-dimensional constitutive parameters}\label{constpar}

 Our dimensional reduction produces one-dimensional quantities describing the thermo-elastic behaviour of a beam and the dissipation inequality, restricting the constitutive assumptions. At this point, a one-dimensional free energy has to be selected and the work is done. On the other hand, in the spirit of our approach, one may ask if this choice has a three-dimensional counterpart, namely if  constants entering the free energy ($s_E,m_o,c_o$) are related to three-dimensional quantities. One way to accomplish this \textit{parameter identification} consists in equating the one-dimensional energy to the corresponding three-dimensional one \cite{FMPG}.     

The three-dimensional constitutive equation for an isotropic linearly thermo-elastic material (cf. \cite{Carl}) is :
\begin{equation}
\Sb=2\mu\Eb +\lambda (\tr\Eb)\Ib+m(\vartheta-\bar{\vartheta})\Ib,
\end{equation}
whose inverse is:
\begin{equation}
\Eb=\frac{1}{2\mu}\left(\Sb-\frac{\lambda}{3\lambda+2\mu}(\tr\Sb)\Ib  \right)+a(\vartheta-\bar{\vartheta})\Ib,
\end{equation}
where
$$
a:=-\frac{m}{3\lambda+2\mu}
$$
is the \textit{coefficient of thermal expansion} and $\lambda$ and $\mu$ are the Lam\'e coefficients. The deformation tensor $\Eb$ then splits into two parts $\Eb=\Eb_m+\Eb_t$, where
$$
\Eb_m:=\frac{1}{2\mu}\left(\Sb-\frac{\lambda}{3\lambda+2\mu}(\tr\Sb)\Ib  \right), \qquad \Eb_t:=a(\vartheta-\bar{\vartheta})\Ib
$$
are the purely mechanical and purely thermal parts, respectively.
When expressed in terms of stress components, the stored- energy density per unit volume is
\begin{equation}
w^{3D}(\Sb)=w_m^{3D}(\Sb)+w_t^{3D}(\Sb), 
\end{equation}
with
\begin{equation}\label{3den}
\begin{aligned}
& w_m^{3D}(\Sb):=\frac{1}{2}\Sb\cdot\Eb_m=\frac{1}{4\mu}\left(|\Sb|^2-\frac{\lambda}{3\lambda+2\mu}(\tr\Sb)^2  \right),\\
& w_t^{3D}(\Sb):=\frac{1}{2}\Sb\cdot\Eb_t=\frac{1}{2}a(\vartheta-\bar{\vartheta})\tr\Sb.
\end{aligned}
\end{equation}
Their one-dimensional counterparts, on following \eqref{constase1}, are
\begin{equation}
w_m^{1D}(N)=\frac{1}{2}\frac{N^2}{s_E}, \qquad w_t^{1D}(N)=\frac{1}{2}\delta_o\, N(\vartheta_o-\bar{\vartheta}_o).
\end{equation}
For a purely axial problem $\Sb=(\Sb\cdot\zb\otimes\zb)\,\zb\otimes\zb=:S_{zz}\,\zb\otimes\zb$, and then \eqref{3den} reduces to:
\begin{equation}
\begin{aligned}
& w_m^{3D}=\frac{1}{E}S_{zz}^2, \quad w_t^{3D}=\frac{1}{2}\,a(\vartheta-\bar{\vartheta})S_{zz}, 
\end{aligned}
\end{equation}
where 
$$
E:=\frac{\mu(3\lambda+2\mu)}{\lambda+2\mu}
$$ 
is the Young modulus. Following \cite{FMPG},  we firstly identify  the extensional stiffness $s_E$ by imposing that
\begin{equation}
\frac{1}{2}\frac{N^2}{s_E}\left(=w_m^{1D}=\int_{\Sc}w_m^{3D}  \right)=\frac{1}{2E}\int_\Sc S_{zz}^2,
\end{equation}
hence,
\begin{equation}
s_E:=E\,\frac{N^2}{\int_\Sc S_{zz}^2}.
\end{equation}
Next, since the factor multiplying the Young modulus $E$ has the dimension of an area, we give $s_E$ the following form:
\begin{equation}
s_E=\frac{EA}{\chi_e}, \quad \chi_e:=A\,\frac{\int_\Sc S_{zz}^2}{N^2}, \quad A:=\int_\Sc dA.
\end{equation}

Analogously, we identify the stress-temperature modulus $m_o$ by imposing that
\begin{equation}
\frac{1}{2}\delta_o\, N(\vartheta_o-\bar{\vartheta}_o)\left(=w_t^{1D}=\int_{\Sc}w_t^{3D}  \right)=\frac{1}{2}a\int_\Sc(\vartheta-\bar{\vartheta})\,S_{zz},
\end{equation}
hence,
\begin{equation}
\delta_o:=a\,\frac{\int(\vartheta-\bar{\vartheta})\,S_{zz}}{N(\vartheta_o-\bar{\vartheta}_o)},
\end{equation}
that we rewrite as:
\begin{equation}
\delta_o=\frac{aA}{\chi_t}, \quad \chi_t:=A\, \frac{N(\vartheta_o-\bar{\vartheta}_o)}{\int(\vartheta-\bar{\vartheta})\,S_{zz}}.
\end{equation}
Note that  `mechanical shape factor'  $\chi_e$ equals 1 whatever the shape of the cross section $\Sc$ if  the field $S_{zz}$ is constant-valued  over $\Sc$, just as it happens to be in Saint-Venant's case of normal force; analogously, the `thermal shape factor' $\chi_t=1$ if there is a constant temperature over the cross section and  the field $S_{zz}$ is constant-valued or linear in $x_\beta$.
In order to achieve a similar identification for the heat capacity, let us consider a rigid three-dimensional conductor,  whose classical free energy is
\begin{equation}
\psi^{3D}=-c\vartheta\log\vartheta,
\end{equation}
where $c$ is the heat capacity; the corresponding one-dimensional energy we have picked is
\begin{equation}
\psi^{1D}=-c_o\vartheta_o\log\vartheta_o.
\end{equation}
We conclude that 
\begin{equation}
c_o=\frac{cA}{\chi_c}, \quad \chi_c:=A\,\frac{\vartheta_o\log\vartheta_o}{\int_{\Sc}\vartheta\log\vartheta};
\end{equation}
note that $\chi_c=1$ if there is a constant temperature over the cross section.

\section{A second example:  bending vibrations}
In this section we  assume that $\hb=h_o\zb$; moreover, we restrict our attention to plane beams, laying in the plane spanned by $\zb$ and $\cb_1$, so that $x-o=x_1\cb_1$. In order to take into account just bending effects, we consider  the following displacement \textit{Ans\"atze}:
\begin{equation}
\begin{aligned}
& \ub(x,\zeta,t)=v(\zeta,t)\cb_1-x_1v'(\zeta,t)\cb_1,\\
& \alpha(x,\zeta,t)=(x-o)\cdot\alphab_1=x_1\alpha_1, \quad \alpha_1:=\alphab_1\cdot\cb_1.
\end{aligned}
\end{equation}

With this assumptions, it is not difficult to see that the pure (without the reactive part) balance equations are
\begin{equation}
\left\{ \begin{array}{ll}
M''=-\rho_o\ddot{v}\\[.2cm]
\dot{\eta}_1=h_1'-\xi_1,
\end{array} \right.
\end{equation}
where we set
$$
\eta_1:=\etab_1\cdot\cb_1, \qquad \xi_1:=\xib_1\cdot\cb_1.
$$
The dissipation inequality \eqref{diss}$_2$ becomes:
\begin{equation}\label{dis1}
\dot{\psi}_1\leq-\eta_1\dot\vartheta_1-h_1\vartheta_1'+M\dot \kappa, 
\end{equation}
with $M:=\mb\cdot\cb_1$ the bending moment, $\vartheta_1= \vthetab_1\cdot\cb_1$ and $\kappa:=-v''$ the curvature. The quantities in need of a constitutive prescription depend on the same list of variables:
\begin{equation}
\begin{aligned}
&\psi_1=\widehat{\psi}_1(\kappa,\vartheta_1,\vartheta_1'), \quad \eta_1=\widehat{\eta}_1(\kappa,\vartheta_1,\vartheta_1'), \\ & M=\widehat{M}(\kappa,\vartheta_1,\vartheta_1').
\end{aligned}
\end{equation}
The Coleman-Noll procedure yields:
\begin{equation}\label{constase11}
\begin{aligned}
&\widehat{\psi}_1 \;{\rm is \; independent \; of \; \vartheta_1'}, \quad \widehat{\eta}_1(\kappa,\vartheta_1)=-\partial_{\vartheta_1}\widehat{\psi}_1(\kappa,\vartheta_1), \\ & \widehat{M}(\kappa,\vartheta_1)=\partial_\kappa\widehat{\psi}_1(\kappa,\vartheta_1),
\end{aligned}
\end{equation}
and moreover, 
$$
\widehat{h}_1(\kappa,\vartheta_1,\vartheta_1')\,\vartheta_1'\leq 0.
$$
We now choose the following free energy:
\begin{equation}
\widehat{\psi}_1(\kappa,\vartheta_1)=\frac{1}{2}s_B\,\kappa^2+m_1\,\kappa(\vartheta_1-\bar{\vartheta}_1)-c_1\vartheta_1\log\vartheta_1,
\end{equation}
where $s_B$ is the \textit{bending stiffness},  $m_1=-\delta_o\,s_B$ the \textit{stress-temperature modulus}, $\vartheta_1$ a prescribed (constant) value of the deviational temperature, and $c_1$ the heat capacity. If we neglect the dissipation and make the small perturbation assumption $\vartheta_1=\widetilde{\vartheta}_1+\theta_1(\zeta,t)$, as done in the previous section, we obtain the following balance equations:

\begin{equation}
\left\{ \begin{array}{ll}
v''''+\delta_o\,\theta_1''=-\dfrac{\rho}{s_B}\ddot{v},\\[.2cm]
c_1\dot{\theta}-\chi_1\theta_1''=\widetilde{\vartheta}_1\,\delta\, s_B\,\dot{v}''\,.
\end{array} \right.
\end{equation}

\vspace{0.3cm}
\remark
An easy computation, analogous to the one carried out in the previous section, shows that
\begin{equation}
\begin{aligned}
& s_B=\frac{EJ}{\chi_b}, \quad \chi_b:=J\,\frac{\int_{\Sc} S_{zz}^2}{M^2}, \\
& c_1=\frac{cA}{\chi_{c1}}, \quad \chi_{c_1}:=A\,\frac{\vartheta_1\log\vartheta_1}{\int_{\Sc}\vartheta\log\vartheta}
\end{aligned}
\end{equation}
for 
\begin{equation}
M=\int_{\Sc}x_1 S_{zz}, \quad J=\int_\Sc x_1^2;
\end{equation}
note that $\chi_b=1$ if $S_{zz}$ is constant over $\Sc $, just it happens to be for $\chi_e$ and $\chi_{c1}=1$ if the temperature over the cross section has null average.

\section*{Acknowledgements}
I thank Prof. Paolo Podio-Guidugli for the time he spent in reading my manuscript and for his many fruitful advice.


\begin{thebibliography}{10}

\bibitem{ABE}
H.~Altenbach, M.~B\^irsan, and V.A. Eremeyev.
\newblock On a thermodynamic theory of rods with two temperature fields.
\newblock {\em Acta Mech.}, 223:1583--1596, 2012.

{\color{black}\bibitem{al}J.J. Alibert, P. Seppecher and F. dell'Isola. Truss modular beam with deformation energy depending on higher displacement gradients, \textit{Math. Mech. Sol.},  8 (1), 51-73,  2003.}

{\color{black} \bibitem{Auff} N. Auffray, F. dell'Isola, V. Eremeyev, A. Madeo and G. Rosi. Analytical continuum mechanics à la Hamilton-Piola: least action principle for second gradient continua and capillary fluids, \textit{Math. Mech. Sol.}, to appear.}

\bibitem{BFPG1}
S.~Bargmann, A.~Favata, and P.~Podio-Guidugli.
\newblock On energy and entropy influxes in the
  \uppercase{G}reen-\uppercase{N}aghdi type \uppercase{III} theory of heat
  conduction.
\newblock {\em Proc. R. Soc. Lond. A}, 469(2152), 2013.

\bibitem{BFPG}
S.~Bargmann, A.~Favata, and P.~Podio-Guidugli.
\newblock A revised exposition of the \uppercase{G}reen-\uppercase{N}aghdi
  theory of heat propagation.
\newblock {\em J. Elast.},  114(2): 143--154, 2014.

{\color{black} \bibitem{Bir} M. B\^irsan and H. Altenbach. Theory of thin thermoelastic rods made of porous
materials. \textit{Arch. Appl. Mech.} 81, 1365--1391,  2011.
}

\bibitem{Carl}
D.E. Carlson.
\newblock Linear thermoelasticity.
\newblock In C.~Truesdell, editor, {\em Encyclopedia of Physics}, volume VI a/2
  of {\em Ed. Fl\"ugge}, pages 297--345. Springer Berlin, 1972.

\bibitem{CN}
B.~Coleman and W.~Noll.
\newblock The thermodynamics of elastic materials with heat conduction and
  viscosity.
\newblock {\em Arch. Rational Mech. Anal}, 13:167--178, 1963.

\bibitem{CBF}
P.~Colli, E.~Bonetti, and M.~Fr\'emond.
\newblock Entropy balance versus energy balance. \uppercase{A}pplication to the
  heat equation and to phase transitions.
\newblock In M.~Fr\'emond and F.~Maceri, Eds., {\em Mechanical Modelling and
  Computational Issues in Civil Engineering}, volume~23 of {\em Lecture Notes
  in Applied and Computational Mechanics}, pages 379--388. Springer Berlin
  Heidelberg, 2005.
  
  {\color{black}\bibitem{dells1} F. dell'Isola and P. Seppecher. Edge Contact Forces and Quasi-Balanced Power, \textit{Meccanica}, 32,  33--52, 1997.}
  
  {\color{black} \bibitem{dellis} F. dell'Isola, U. Andreaus and L. Placidi. At the origins and in the vanguard of peri-dynamics, non-local and higher gradient continuum mechanics. An underestimated and still topical contribution of Gabrio Piola, \textit{Math. Mech. Sol.}, to appear.}

\bibitem{DSW}
C.N. DeSilva and A.B. Whitman.
\newblock Thermodynamical theory of directed curves.
\newblock {\em J. Math. Phys.}, 12:1603--1609, 1971.

\bibitem{FMPG}
A.~Favata, A.~Micheletti, and P.~Podio-Guidugli.
\newblock On shear and torsion factors in the theory of linearly elastic rods.
\newblock {\em J. Elast.}, 99(2):203--210, 2010.

\bibitem{FPG1}
A.~Favata and A.~Podio-Guidugli.
\newblock What shell theory fits carbon nanotubes?
\newblock {\em Adv. Struct. Mat.}, 15, 561--570, 2011.

\bibitem{FPG}
A.~Favata and A.~Podio-Guidugli.
\newblock A new \uppercase{CNT}-oriented shell theory.
\newblock {\em Europ. J. Mech. A/Solids}, 35:75--96, 2012.

\bibitem{FPGT}
A.~Favata, P.~Podio-Guidugli, and G.~Tomassetti.
\newblock Energy splitting theorems for materials with memory.
\newblock {\em J. Elast.}, 101(1):59--67, 2010.

\bibitem{Fr}
M.~Fr\'emond.
\newblock The basic laws of mechanics.
\newblock In {\em Phase Change in Mechanics}, volume~13 of {\em Lecture Notes
  of the Unione Matematica Italiana}, pp. 5--29. Springer Berlin Heidelberg,
  2012.

\bibitem{Ger}
P.~Germain.
\newblock La m\'ethode des puissances virtuelles en m\'ecanique des milieux
  continus, premi\`ere partie: th\'eorie du second gradient.
\newblock {\em J. M\'ecanique}, 12:235--274, 1973.

\bibitem{GN}
A.E. Green and P.M. Naghdi.
\newblock Non-isothermal theory of rods, plates and shells.
\newblock {\em Int. J. Solids Struct.}, 6:209--244, 1970.

\bibitem{GN_1977}
A.E. Green and P.M. Naghdi.
\newblock On thermodynamics and the nature of the second law.
\newblock {\em Proc. R. Soc. Lond. A}, 4:253--270, 1977.

\bibitem{GN1}
A.E. Green and P.M. Naghdi.
\newblock On thermal effects in the theory of rods.
\newblock {\em Int. J. Solids Struct.}, 15:829--853, 1979.

\bibitem{GN_1991}
A.E. Green and P.M. Naghdi.
\newblock demonstration of consistency of an entropy balance with balance of
  energy.
\newblock {\em J. Appl. Math. Phys. (ZAMP)}, 42:159--168, 1991.

\bibitem{GN_1992}
A.E. Green and P.M. Naghdi.
\newblock On undamped heat waves in an elastic solid.
\newblock {\em J. Thermal Stresses}, 15:253--264, 1992.

{\color{black} \bibitem{GR} A.E. Green, R.S. Rivlin. Multipolar continuum mechanics. \textit{Arch. Rat.
Mec. Anal.}, 17, 113--147, 1964.}

{\color{black} \bibitem{GR1} A.E. Green, R.S. Rivlin. Multipolar continuum mechanics: functional theory. I.
\textit{Proc. R. Soc. Lond.  A}, 284, 303--324, 1965.}

{\color{black} \bibitem{GR2} A.E. Green, R.S. Rivlin.  On Cauchy's equations of motion. \textit{ZAMP}, 15, 290--292, 1964.}

{\color{black} \bibitem{GR3} A.E. Green, R.S. Rivlin. Simple force and stress multipoles. Arch. Rat.
Mec. Anal., 16, 325--353, 1964.}

{\color{black}\bibitem{Ies} D. Ie\c{s}an. Thermal effects in orthotropic porous elastic beams, \textit{ZAMP}, 60, 138--153,  2009.}

\bibitem{Jones_1966}
J.P. Jones.
\newblock Thermoelastic vibrations of a beam.
\newblock {\em J. Acoust. Soc. Am.}, 39:542--548, 1966.

%

\bibitem{K}
C.B. Kafadar.
\newblock On the nonlinear theory of rods.
\newblock {\em Int. J. Eng. Sci.}, 10:369--391, 1972.

\bibitem{LPG}
M.~Lembo and P.~Podio-Guidugli.
\newblock Internal constraints, reactive stresses, and the
  \uppercase{T}imoshenko beam theory.
\newblock {\em J. Elast.}, 65:131--148, 2001.

\bibitem{Mau_1980}
G.A. Maugin.
\newblock The method of virtual power in continuum mechanics: Application to
  coupled fields.
\newblock {\em Acta Mech.}, 35(1-2):1--70, 1980.

\bibitem{Mue_1971}
I.~M\"uller.
\newblock The coldness, a universal function in thermoelastic bodies.
\newblock {\em Arch. Rat. Mech. Anal.}, 41(5):319--332,
  1971.

{\color{black}\bibitem{pid} C. Pideri and P. Seppecher. A second gradient material resulting from the homogenization of an heterogeneous linear elastic medium,\textit{ Cont. Mech. Thermod.}, 9 (5), 241-257,  1997. }

{\color{black} \bibitem{Pio} G. Piola. \textit{Memoria intorno alle equazioni fondamentali del movimento di corpi
qualsivogliono considerati secondo la naturale loro forma e costituzione}- Modena,
Tipi del R.D. Camera, 1845-1846.}

\bibitem{PPG_1989}
P.~Podio-Guidugli.
\newblock An exact derivation of the thin plate equation.
\newblock {\em J. Elast.}, 22(2-3):121--133, 1989.

\bibitem{PPG_1997}
P.~Podio-Guidugli.
\newblock Inertia and invariance.
\newblock {\em Ann. Mat. Pur. Appl.}, 172:103--124, 1997.

\bibitem{PPG_2008}
P.~Podio-Guidugli.
\newblock Concepts in the mechanics of thin structures.
\newblock In A.~Morassi and R.~Paroni, Eds., {\em Classical and Advanced
  Theories of Thin Structures}, Vol. 503 of {\em CISM International Centre
  for Mechanical Sciences}, pp. 77--109. Springer Vienna, 2008.

\bibitem{PPG}
P.~Podio-Guidugli.
\newblock A virtual power format for thermomechanics.
\newblock {\em Cont. Mech. Thermodyn.}, 20:479--487, 2009.

\bibitem{PPG_2013}
P.~Podio-Guidugli.
\newblock Dissipative entropy makes the heat equation hyperbolic.
\newblock {\em Atti della Accademia Peloritana dei Pericolanti}, 91, 2013.

\bibitem{primer2}
P.~Podio-Guidugli.
\newblock {\em A primer in Elasticity (second edition)}.
\newblock Springer Netherlands, Forthcoming.

\bibitem{Sim}
J.G. Simmonds.
\newblock A simple nonlinear thermodynamic theory of arbitrary elastic beams.
\newblock {\em J. Elast.}, 81:51--62, 2005.


{\color{black}
\bibitem{Tim} S.P. Timoshenko.
On the correction factor for shear of the differential equation for transverse vibrations of bars of uniform cross-section, \textit{Phil. Mag.}, 744, 1921.

\bibitem{Tim1} S.P. Timoshenko. On the transverse vibrations of bars of uniform cross-section, \textit{Phil. Mag.}, 125, 1922.}

\bibitem{Tr}
C.~Truesdell and W.~Noll.
\newblock The non-linear field theory of mechanics.
\newblock In S.~Fl\"ugge, Ed., {\em Encyclopedia of Physics}, volume III/3.
  Springer New York, 1965.

\end{thebibliography}
\end{document}